\newcommand{\CLASSINPUTtoptextmargin}{2cm}
\newcommand{\CLASSINPUTbottomtextmargin}{2cm}
\documentclass[conference, a4paper]{IEEEtran}

\ifCLASSINFOpdf
  \usepackage[pdftex]{graphicx}
\else
\fi
\usepackage{amsmath}
\usepackage{amssymb}
\usepackage{algorithmic}
\usepackage{algorithm}
\ifCLASSOPTIONcompsoc
  \usepackage[caption=false,font=normalsize,labelfont=sf,textfont=sf]{subfig}
\else
  \usepackage[caption=false,font=footnotesize]{subfig}
\fi
\usepackage{booktabs}
\usepackage{multirow}
\usepackage{url}
\usepackage{balance}
\usepackage{comment}
\usepackage{mathtools}

\usepackage{color}

\def\sub#1{_{\rm #1}}

\def\eg{{\it e.g.}}

\def\ie{{\it i.e.}}

\DeclareMathOperator*{\argmax}{arg\,max}
\usepackage[short]{optidef}

\begin{document}
%
\title{Client Selection for Federated Learning with \\Heterogeneous Resources in Mobile Edge}

\author{\IEEEauthorblockN{Takayuki Nishio}
\IEEEauthorblockA{Graduate School of Informatics,\\
Kyoto University, Japan\\
Email: nishio@i.kyoto-u.ac.jp}
\and
\IEEEauthorblockN{Ryo Yonetani}
\IEEEauthorblockA{Institute of Industrial Science,\\The University of Tokyo, Japan\\
Email: yonetani@iis.u-tokyo.ac.jp}}


%


\maketitle


\begin{abstract}
We envision a mobile edge computing (MEC) framework for machine learning (ML) technologies, which leverages distributed client data and computation resources for training high-performance ML models while preserving client privacy. 
Toward this future goal, this work aims to extend Federated Learning (FL), a decentralized learning framework that enables privacy-preserving training of models, to work with heterogeneous clients in a practical cellular network. 
The FL protocol iteratively asks random clients to download a trainable model from a server, update it with own data, and upload the updated model to the server, while asking the server to aggregate multiple client updates to further improve the model. While clients in this protocol are free from disclosing own private data, the overall training process can become inefficient when some clients are with limited computational resources (\ie, requiring longer update time) or under poor wireless channel conditions (longer upload time). 
Our new FL protocol, which we refer to as \texttt{FedCS}, mitigates this problem and performs FL efficiently while actively managing clients based on their resource conditions. Specifically, \texttt{FedCS} solves a client selection problem with resource constraints, which allows the server to aggregate as many client updates as possible and to accelerate performance improvement in ML models. We conducted an experimental evaluation using publicly-available large-scale image datasets to train deep neural networks on MEC environment simulations. The experimental results show that \texttt{FedCS} is able to complete its training process in a significantly shorter time compared to the original FL protocol.
\end{abstract}


%
\IEEEpeerreviewmaketitle

\section{Introduction}\label{sec:intro}
A variety of modern AI products are powered by cutting-edge machine learning (ML) technologies, which range from face detection and language translation installed on smartphones to voice recognition and speech synthesis used in virtual assistants such as Amazon Alexa and Google Home. Therefore, the development of such AI products typically necessitates large-scale data, which are essential for training high-performance ML models such as a deep neural network. Arguably, a massive amount of IoT devices, smartphones, and autonomous vehicles with high-resolution sensors, all of which are connected to a high-speed network, can serve as promising data collection infrastructure in the near future (\eg,~\cite{Choi2016}). Researchers in the field of communication and mobile computing have started to interact with data science communities in the last decade and have proposed mobile edge computing (MEC) frameworks that can be used for large-scale data collection and processing~\cite{Hu2015}.

Typically, MEC frameworks assume that all data resources are transferred from data collection clients (IoT devices, smartphones, and connected vehicles) to computational infrastructure (high-performance servers) through cellular networks to perform their tasks~\cite{Liu2017,Kato2017}.
However, this assumption is not always acceptable when private human activity data are collected, such as life-logging videos, a history of e-mail conversations, and recorded phone calls. On one hand, such private activity data would be a key factor for improving the quality of AI products that support our daily life, which include not only AI-related apps on smartphones and virtual assistants but also AI-powered smart cities. On the other hand, uploading these data directly to computational infrastructure is problematic as the data could be eavesdropped by malicious users in a network to compromise client's privacy.

\begin{figure}[t]
  \begin{center}
    \includegraphics[width=\linewidth]{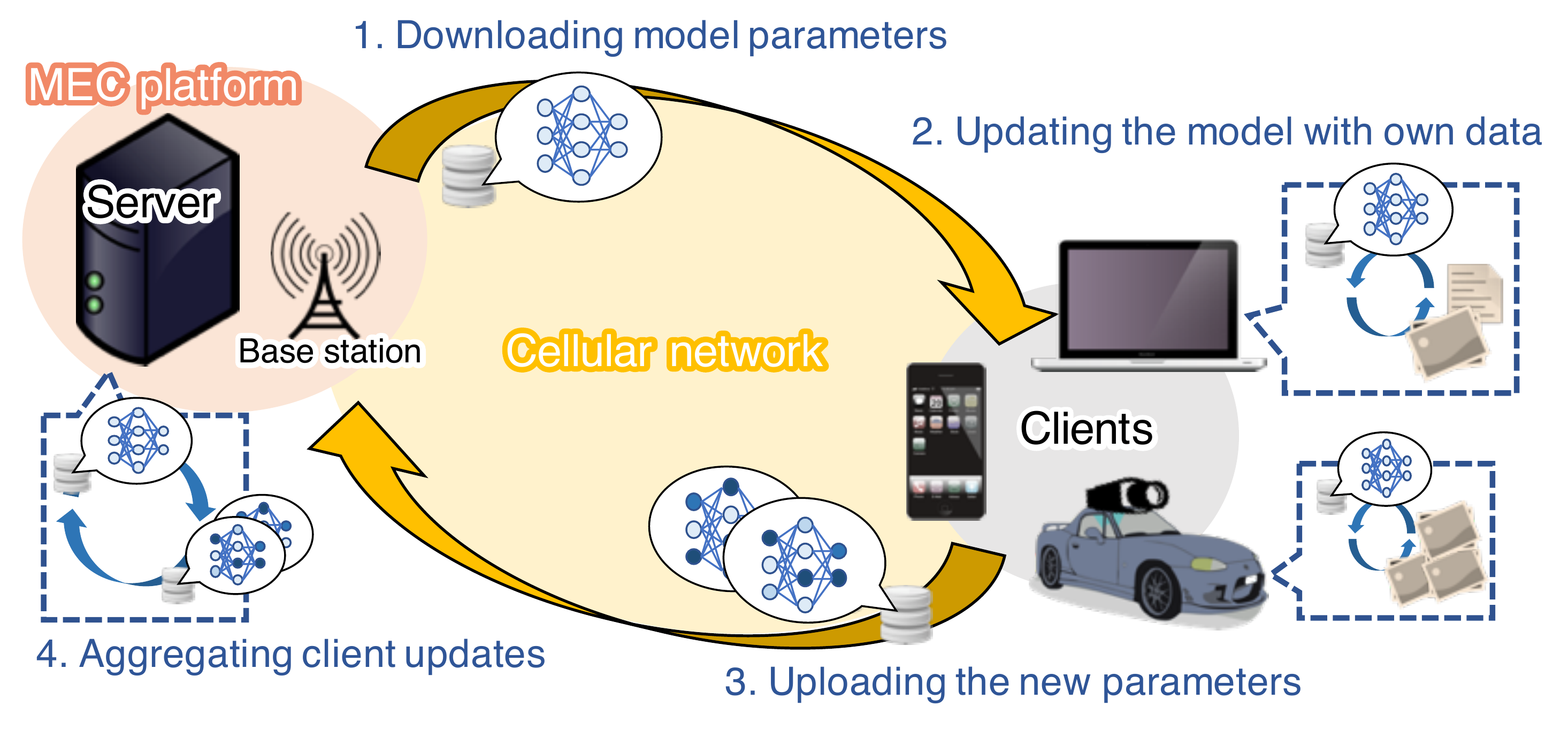}
    \caption{{\bf Federated learning}~\cite{Mcmahan2017} enables one to train machine learning models on private client data through the iterative communications of model parameters between a server and clients. How can we implement this training process in practical cellular networks with heterogeneous clients?}
    \label{fig:teaser}
  \end{center}
\end{figure}

To address this fundamental privacy concern, one work has recently been presented by the ML community: \emph{Federated Learning (FL)}~\cite{Mcmahan2017}. As illustrated in Fig.~\ref{fig:teaser}, FL iteratively asks random clients to 1) download parameters of a trainable model from a certain server, 2) update the model with their own data, and 3) upload the new model parameters to the server, while asking the server to 4) aggregate multiple client updates to further improve the model. In exchange for requiring data collection clients to install a certain level of computational resources (\eg, a laptop equipped with reasonable GPUs, autonomous vehicles with moderate computational capacities~\cite{Choi2016}), the FL protocol allows the clients to keep their data secure in their local storage. 

In this work, we focus on the implementation of the abovementioned FL protocol in practical MEC frameworks. We believe that our work will influence the future development platform of various AI products that require a large amount of private activity data to train ML models. In particular, we consider the problem of running FL in a cellular network used by heterogeneous mobile devices with different data resources, computational capabilities, and wireless channel conditions. Unfortunately, a direct application of existing FL protocols without any consideration of such heterogeneous client properties will make the overall training process inefficient. For instance, when some clients are with limited computational resources, they will require longer time to update models. Moreover, if the clients are under poor wireless channel conditions, that will result in longer update time. All such problems will delay the subsequent server's aggregation step necessary to continue the training process.

Our main contribution is a new protocol referred to as \texttt{FedCS}, which can run FL efficiently while an operator of MEC frameworks actively manages the resources of heterogeneous clients. Specifically, \texttt{FedCS} sets a certain deadline for clients to download, update, and upload ML models in the FL protocol. Then, the MEC operator selects clients such that the server can aggregate as many client updates as possible in limited time frames, which makes the overall training process efficient and reduces a required time for training ML models. This is technically formulated by a client-selection problem that determines which clients participate in the training process and when each client has to complete the process while considering the computation and communication resource constraints imposed by the client, which we can solve in a greedy fashion.

We evaluate our approach with a realistic large-scale training of deep neural networks for object classification on a simulated MEC environment, where client data were generated using publicly-available large-scale image datasets. Our experimental results reveal that the \texttt{FedCS} can complete its training process in a significantly shorter time compared to the original FL protocol.

\subsection*{Related Work}
Resource optimization for MEC frameworks is one of the common topics in the field of communication and mobile computing. Recent work includes the joint optimization of heterogeneous data, computation, and communication resources~\cite{Nishio2013,Sardellitti2015,Yu2016}. However, these approaches are designed to minimize computation times and/or energy consumptions for general computation tasks, which is considerably different from our work that aims to maximize the efficiency of training ML models. Moreover, as we stated earlier, our work assumes a different scenario where each mobile client has data and computational resources to preserve client data privacy when performing ML tasks. These differences motivate us to propose new tailored MEC protocols and algorithms.

Federated Learning is an emerging technique in the ML community. Following pioneering work~\cite{Mcmahan2017}, recent studies have specifically focused on how to enhance the security of FL protocols~\cite{Bonawitz2017,Geyer2017}. However, little work has examined how to run FL efficiently with a practical network configuration. One exception is \cite{Konecny2016}, which explored model compression techniques for efficient communications while sacrificing model performances. The other one is \cite{Wang2018}, which optimized hyper-parameters of FL (\ie the number of epochs in each update phase and the number of total epochs) in a resource constrained MEC environment. However, these techniques do not particularly consider heterogeneous computation and communications and/or data resources of clients. The additional use of model compression techniques could help us improve the overall efficiency of our protocol, which is however beyond the scope of this study.


\section{Federated Learning}
In this section, we briefly introduce the original FL framework presented in \cite{Mcmahan2017}. Then, we identify the problems that affect FL communications when they are performed by heterogeneous clients in resource-constrained cellular networks.
\subsection{Federated Learning}
Consider a scenario where a large population of mobile clients individually have data that they want to maintain as secret, such as laptops with personal collections of photos and autonomous vehicles with cityscape images captured by cameras. If all these distributed data are accessible, one can obtain a high-performance ML model that has been trained on an extremely large data collection. However, it is not desirable for clients to disclose their data owing to privacy concerns.

Federated Learning~\cite{Mcmahan2017} is a \emph{decentralized} learning protocol that aims to resolve the abovementioned problem. As shown in Protocol~\ref{prt:FL}, FL asks a certain server and $\lceil K\times C \rceil$ random clients (where $K$ is the number of all clients, $C$ is the fraction of clients considered in each round, and $\lceil \cdot \rceil$ is the ceiling function, ) to communicate the parameters of a global model that they are going to train (\texttt{Distribution} and \texttt{Update and Upload} steps). The protocol requires the selected clients to compute an update of the model using their data (\texttt{Update and Upload} step), while asking the server to aggregate multiple updates from the clients to make the model better (\texttt{Aggregation} step). The advantage of this protocol is that clients do not have to upload private data; instead, they secure the data in their local storage. The only technical requirement is that each client must have a certain level of computational resources because \texttt{Update and Upload} consists of multiple iterations of the forward propagation and backpropagation of the model (\ie, we focus exclusively on training deep neural networks in a supervised manner; see \cite{Mcmahan2017} for more details). 

\begin{algorithm}[t]
\floatname{algorithm}{Protocol}
\caption{Federated Learning. $K$ is the number of clients that participate in the protocol. $C\in(0, 1]$ is a hyperparameter that controls the fraction of clients considered in each round.}
\label{prt:FL}
\begin{algorithmic}[1]
\STATE \texttt{Initialization:} The server first initializes a global model randomly or by pretraining with public data.
\STATE \texttt{Client Selection:} The server randomly selects $\lceil K\times C\rceil$ clients.
\STATE \texttt{Distribution:} The server distributes the parameters of the global model to the selected clients.
\STATE \texttt{Update and Upload:} Each selected client updates the global model using their data and uploads the updated model parameters to the server.
\STATE \texttt{Aggregation:} The server averages the updated parameters and replaces the global model by the averaged model.
\STATE All steps but \texttt{Initialization} are iterated until the global model achieves a desired performance.
\end{algorithmic}
\end{algorithm}

\subsection{Heterogeneous Client Problem in FL}
Protocol~\ref{prt:FL} can experience major problems while training ML models in a practical cellular network, which are mainly due to the lack of consideration of the heterogeneous data sizes, computational capacities, and channel conditions of each client.
For example, if a client has more data compared to others, the client will require longer time to update models unless it has a better computational resource. This will delay the subsequent communication for uploading new model parameters. Moreover, upload time will be longer if a client is under a severely poor channel condition.

All such problems about heterogeneous client resources will become bottlenecks in the FL training process; the server can complete the \texttt{Aggregation} step \emph{only after it receives all client updates}. One may set a deadline for random clients to complete the \texttt{Update and Upload} step and ignore any update submitted after the deadline. However, this straightforward approach will lead to the inefficient use of network bandwidths and waste the resources of delayed clients.

\section{\texttt{FedCS}: Federated Learning with \\Client Selection}

We propose a new FL protocol, \texttt{FedCS}, which works efficiently with clients with heterogeneous resources. In the following sections, we first summarize several assumptions of our proposal and then present \texttt{FedCS} in more detail.

\subsection{Assumptions}\label{sec:assumption}

As illustrated in Fig.~\ref{fig:teaser}, we consider that a certain MEC platform, which is located in a wireless network and consists of a server and a base station (BS), manages the behaviors of the server and clients in the FL protocol. 
We will particularly focus in this work on leveraging the wireless networks when they are stable and not congested, such as at midnight or in the early morning time, mainly because ML models to be trained and communicated are typically large. Nevertheless, each process has to be carried out under certain limited bandwidths, particularly when there are multiple ML tasks to be performed via FL. Specifically, we assume that the amount of resource blocks (RBs; the smallest unit of bandwidth resources defined in LTE \cite{sesia2011}) available for each process is limited and managed by the MEC operator. In addition, if multiple clients upload model parameters simultaneously, the throughput for each client decreases accordingly.

We assume that the modulation and coding scheme of radio communications for each client are determined appropriately while considering its channel state so that packet-loss rate is negligible. This leads to different throughput for each client to upload model parameters although the amount of allocated RBs is constant. The throughput for broadcast and multicast transmission by the BS is assumed to be limited by that of the client with the worst channel conditions. Nevertheless, we also assume the channel state and throughput of each client to be stable as mentioned above. 

\subsection{\texttt{FedCS} Protocol}
\begin{figure}[t]
  \begin{center}
    \includegraphics[width=\linewidth]{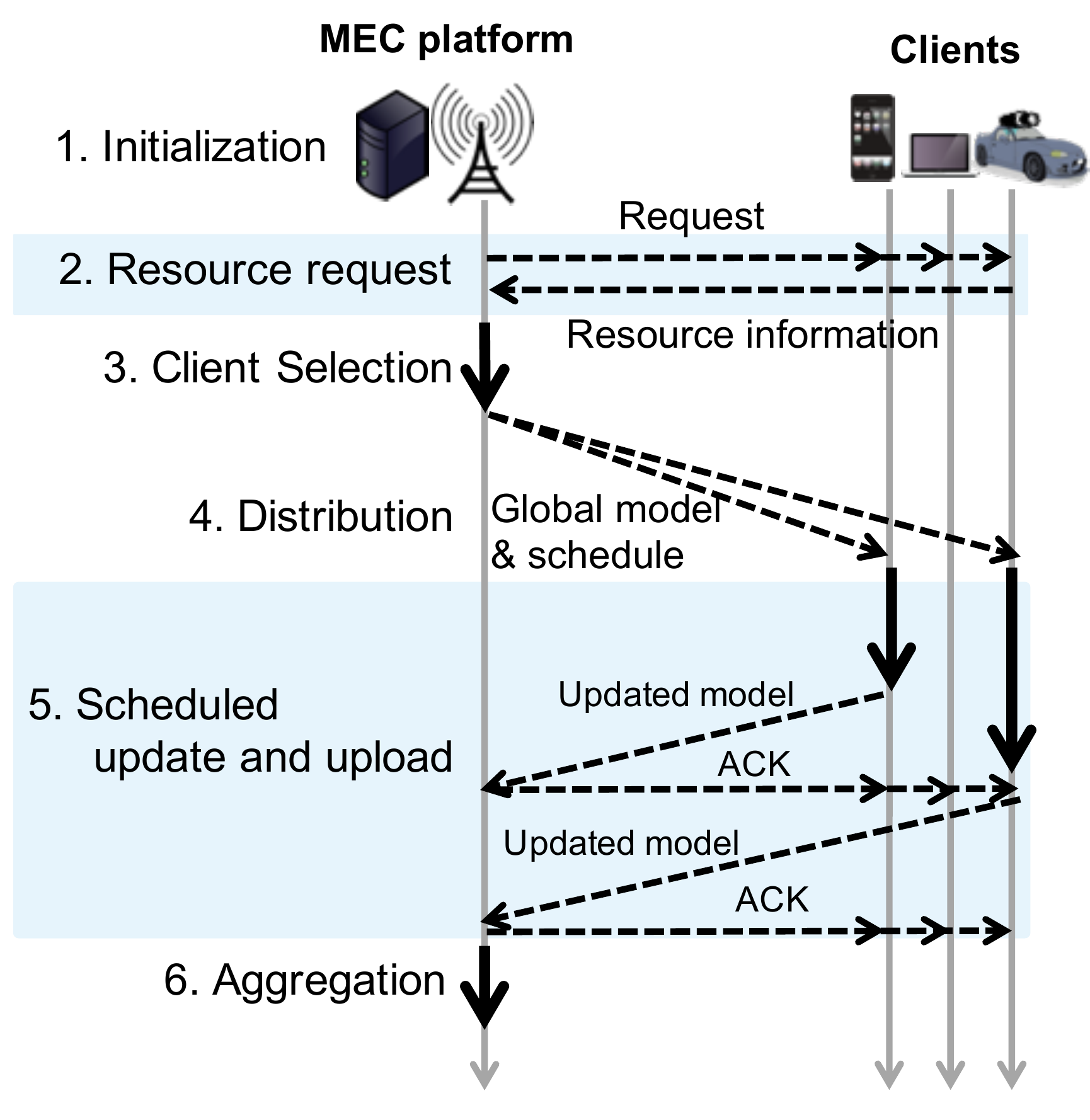}
    \caption{{\bf Overview of \texttt{FedCS} Protocol}. Solid black lines denote computation processes while dashed lines indicate wireless communications.\vspace{-2mm}}
    \label{fig:procedure}
  \end{center}
\end{figure}

We present \texttt{FedCS} in Protocol~\ref{prt:FedCS} (see also the diagram in Fig.~\ref{fig:procedure} for how each step is performed in order). The key idea of our protocol is that instead of selecting random clients in the original \texttt{Client Selection} step of Protocol~\ref{prt:FL}, we propose the following two-step client selection scheme. First, the new \texttt{Resource Request} step asks random clients to inform the MEC operator of their resource information such as wireless channel states, computational capacities (\eg, if they can spare CPUs or GPUs for updating models), 
and the size of data resources relevant to the current training task (\eg, if the server is going to train a `dog-vs-cat' classifier, the number of images containing dogs or cats). Then, the operator refers to this information in the subsequent \texttt{Client Selection} step to estimate the time required for the \texttt{Distribution} and \texttt{Scheduled Update and Upload} steps and to determine which clients go to these steps (the specific algorithms for scheduling clients are explained later).
In the \texttt{Distribution} step, a global model is distributed to the selected clients via multicast from the BS because it is bandwidth effective for transmitting the same content (\ie, the global model) to client populations. In the \texttt{Scheduled Update and Upload} step, the selected clients update the model in parallel and upload new parameters to the server using the RBs allocated by the MEC operator. The server aggregates client updates following Protocol~\ref{prt:FL} and measures model performances with certain validation data. Until the model achieves a certain desired performance (\eg, a classification accuracy of 90\%) or the final deadline arrives, all steps but \texttt{Initialization} are iterated for multiple rounds.

\begin{algorithm}[t]
\floatname{algorithm}{Protocol}
\caption{Federated Learning with Client Selection. $K$ is the number of clients, and $C\in(0, 1]$ describes the fraction of random clients that receive a resource request in each round.}
\label{prt:FedCS}
\begin{algorithmic}[1]
\STATE \texttt{Initialization} in Protocol~\ref{prt:FL}.
\STATE \texttt{Resource Request:} The MEC operator asks $\lceil K\times C \rceil$ random clients to participate in the current training task. Clients who receive the request notify the operator of their resource information.
\STATE \texttt{Client Selection:} Using the information, the MEC operator determines which of the clients go to the subsequent steps to complete the steps within a certain deadline.
\STATE \texttt{Distribution:} The server distributes the parameters of the global model to the selected clients.
\STATE \texttt{Scheduled Update and Upload:} The clients update global models and upload the new parameters using the RBs allocated by the MEC operator.
\STATE \texttt{Aggregation} in Protocol~\ref{prt:FL}.
\STATE All steps but \texttt{Initialization} are iterated for multiple rounds until the global model achieves a desired performance or the final deadline arrives.
\end{algorithmic}
\end{algorithm}

\subsection{Algorithm for \texttt{Client Selection} Step}

Our goal in the \texttt{Client Selection} step is to allow the server to aggregate as many client updates as possible within a specified deadline. This criterion is based on the result from \cite{Mcmahan2017} that a larger fraction of clients used in each round saves the time required for global models to achieve a desired performance. 
Based on the criterion, the MEC operator selects clients who can complete the \texttt{Distribution} and \texttt{Scheduled Update and Upload} steps within a deadline. At the same time, the operator schedules when the RBs for model uploads are allocated to the selected clients to prevent congestion in the limited bandwidths a cellular network could impose. Note that we assume that selected clients start and complete their upload processes one by one for simplicity. Nevertheless, even if multiple clients can upload in parallel by sharing RBs, the time required for transmitting all models is the same as that for the sequential upload.

Formally, let $\mathbb{K}=\{1,\dots,K\}$ be a set of indices that describes $K$ clients and $\mathbb{K}'\subseteq \mathbb{K}$ be a subset of $\mathbb{K}$ randomly selected in the \texttt{Resource Request} step (\ie, $|\mathbb{K}'|=\lceil K\times C \rceil$). 
$\mathbb{S} = [k_1, k_2, ..., k_i, ..., k_{|\mathbb{S}|}]$, where $k_i \in \mathbb{K}',\;|\mathbb{S}|\leq |\mathbb{K}'|$, denotes a sequence of indices of the clients selected in \texttt{Client Selection}, which we aim to optimize. 
In the \texttt{Update and Upload} step, clients sequentially upload their model in the order of $\mathbb{S}$. 
Let $\mathbb{R}_+$ be the set of non-negative real numbers, $T\sub{round}\in\mathbb{R}_+$ be the deadline for each round, $T\sub{final}\in\mathbb{R}_+$ be the final deadline, and $T\sub{cs}\in\mathbb{R}_+$ and $T\sub{agg}\in\mathbb{R}_+$ be the time required for the \texttt{Client Selection} and \texttt{Aggregation} steps, respectively. $T_\mathbb{S}^\mathrm{d}\in\mathbb{R}_+$ denotes the time required for \texttt{Distribution}; it depends on selected clients $\mathbb{S}$. $t_k^\mathrm{UD}\in\mathbb{R}_+$ and $t_k^\mathrm{UL}\in\mathbb{R}_+$ denote the time consumed by the $k$-th client to update and upload models, respectively. These client-wise parameters can be determined based on the resource information notified in the \texttt{Resource Request} step. 

Now, the objective of \texttt{Client Selection}, namely accepting as many client updates as possible, can be achieved by maximizing the number of selected clients, \ie, $\max_{\mathbb{S}} |\mathbb{S}|$.
To describe the constraint, we define the estimated elapsed time from the beginning of the \texttt{Scheduled Update and Upload} step until the $k_i$-th client completes the update and upload procedures, as follows:
\begin{equation}
\Theta_i \coloneqq \begin{cases}
		0 & \text{if}\ i = 0; \\
  		T^\mathrm{UD}_i + T^\mathrm{UL}_i & \text{otherwise},
      \end{cases}
\end{equation}
\begin{eqnarray}
        T^\mathrm{UD}_i &=& \sum_{j=1}^{i} \max \{0, t_{k_j}^\mathrm{UD} - \Theta_{j-1}\}, \\
  		T^\mathrm{UL}_i &=& \sum_{j=1}^{i} t_{k_j}^\mathrm{UL}.
\end{eqnarray}
As clients upload their model updates one by one, $T^\mathrm{UL}_i$ is the accumulation of all required upload times, $t^\mathrm{UL}_{k_j}$. In contrast, model updates can be performed while the prior clients are in the upload step. Therefore, individual update times, $t^\mathrm{UD}_{k_j}$, will not consume $T^\mathrm{UD}_i$ as long as they are within the previous elapsed time, $\Theta_{j-1}$.

In summary, \texttt{Client Selection} is formulated by the following maximization problem with respect to $\mathbb{S}$:
\begin{maxi}|s|
 	{\substack{\mathbb{S}}}{|\mathbb{S}|}
 	{\label{maxiprob}}{}
   \addConstraint{T\sub{round} \geq}{T\sub{cs} + T_\mathbb{S}^\mathrm{d} + \Theta_{|\mathbb{S}|} + T\sub{agg}.}{ }
\end{maxi}

{\bf Optimization strategies:}
Solving the maximization problem~\eqref{maxiprob} is nontrivial as it requires a complex combinatorial optimization where the order of elements in $\mathbb{S}$ affects $T_{|\mathbb{S}|}$. To this end, we propose a heuristic algorithm based on the greedy algorithm for a maximization problem with a knapsack constraint~\cite{sviridenko2004}. As shown in Algorithm~\ref{alg:scheduling}, we iteratively add the client that consumes the least time for the model upload and update (steps 3, 4, and 9) to $\mathbb{S}$ until elapsed time $t$ reaches deadline $T\sub{round}$ (steps 5, 6, 7, and 8).
The order of the algorithm is $O(|\mathbb{K}'||\mathbb{S}|)$, which is considerably less than that of a naive brute force search, $O(2^{|\mathbb{K}|'}!)$.

\begin{algorithm}[t]
\floatname{algorithm}{Algorithm}
\caption{\texttt{Client Selection} in Protocol~\ref{prt:FedCS}}
\label{alg:scheduling}
\begin{algorithmic}[1]
\REQUIRE{Index set of randomly selected clients $\mathbb{K}'$}
\STATE \textbf{Initialization} $\mathbb{S} \gets \{\}$, $T_{\mathbb{S}=\emptyset}^\mathrm{d} \gets 0$, $\Theta \gets 0$
\WHILE{$|\mathbb{K}'| > 0$}
 	\STATE{$x \gets \argmax_{k\in \mathbb{K}'} \frac{1}{T_{\mathbb{S}\cup k}^\mathrm{d}-T_{\mathbb{S}}^\mathrm{d} + t_{k}^\mathrm{UL} + \max \{0, t_{k}^\mathrm{UD} - \Theta\}}$}
	\STATE{remove $x$ from $\mathbb{K}'$}
	\STATE{$\Theta' \gets \Theta + t_{x}^\mathrm{UL} + \max \{0, t_{x}^\mathrm{UD} - \Theta\}$}
  	\STATE{$t \gets T\sub{cs} + T_{\mathbb{S} \cup x }^\mathrm{d} + \Theta' + T\sub{agg}$}
      \IF{$t < T\sub{round}$}
			\STATE{$\Theta \gets \Theta'$}
         \STATE{add $x$ to $\mathbb{S}$}
      \ENDIF{}
\ENDWHILE
\RETURN{$\mathbb{S}$}
\end{algorithmic}
\end{algorithm}

{\bf Selection of $T\sub{round}$:}
The important parameter in Algorithm~\ref{alg:scheduling} is $T\sub{round}$. If we set $T\sub{round}$ to be large, we expect more clients to be involved in each round (\ie, larger sets of $\mathbb{S}$). However, this simultaneously reduces the possible number of update aggregations until final deadline $T\sub{final}$. Our experimental evaluation shows how different selections of $T\sub{round}$ affect the final performances of trained models.


\section{Performance Evaluation}
As a proof-of-concept scenario to show how our protocol works effectively, we simulated a MEC environment and conducted experiments of realistic ML tasks using publicly-available large-scale datasets.

\subsection{Simulated Environment}
We simulated a MEC environment implemented on the cellular network of an urban microcell consisting of an edge server, a BS, and $K=1000$ clients, on a single workstation with GPUs. The BS and server were co-located at the center of the cell with a radius of 2 km, and the clients were uniformly distributed in the cell. 

Wireless communications were modeled based on LTE networks with a well-known urban channel model defined in the ITU-R M.2135-1 Micro NLOS model of a hexagonal cell layout~\cite{ITU-R}. Carrier frequency was 2.5 GHz, and the antenna heights of the BS and clients were set to 11 m and 1 m, respectively. The transmission power and antenna gain of the BS and clients were respectively assumed to be 20 dBm and 0 dBi for simplicity. As a practical bandwidth limitation, we assumed that 10 RBs, which corresponded to a bandwidth of 1.8 MHz, were assigned to a client in each time slot of 0.5 ms. We employed a throughput model based on the Shannon capacity with a certain loss used in~\cite{Akdeniz2014} with $\Delta = 1.6$ and $\rho_\mathrm{max}=4.8$. With this setting, the mean and maximum throughputs of client $\theta_k$ were 1.4 Mbit/s and 8.6 Mbit/s, respectively, which are realistic values in LTE networks. We consider the throughput obtained from the abovementioned model as the average throughput of each client and used the throughput to calculate $t_x^\mathrm{UL}$ in \texttt{Client Selection}. As mentioned in Section \ref{sec:assumption}, all of the FL processes were assumed to be performed during the network condition was stable and client devices were likely to be unused and stationary at midnight or in the early morning. This allowed us to regard the average throughput as stable. Nevertheless, to take into account a small variation of short-term throughput at \texttt{Scheduled Update and Upload} that can happen in practice, everytime when clients upload models we sampled the throughput from the Gaussian distribution with the mean and standard deviation given by the average throughput and its $r\%$ value, respectively.


The abovementioned assumptions provide concrete settings for several parameters used in Algorithm~\ref{alg:scheduling}. Let $D\sub{m}$ be the data size of the global model. Then, the time required for uploading models can be calculated as $t_{k}^\mathrm{UL} = D\sub{m} / \theta_k$. The time required for model distribution is simply modeled as $T_{\mathbb{S}}^\mathrm{d} = D\sub{m} / \min_{k \in \mathbb{S}}\{\theta_k\}$. 
In addition, we assumed that the computation capability of the server was sufficiently high to neglect the time consumed by \texttt{Client Selection} and \texttt{Aggregation}; thus, $T\sub{cs}=0$ and $T\sub{agg}=0$.

\subsection{Experimental Setup of ML Tasks}
\begin{figure}[t]
  \begin{center}
    \includegraphics[width=\linewidth]{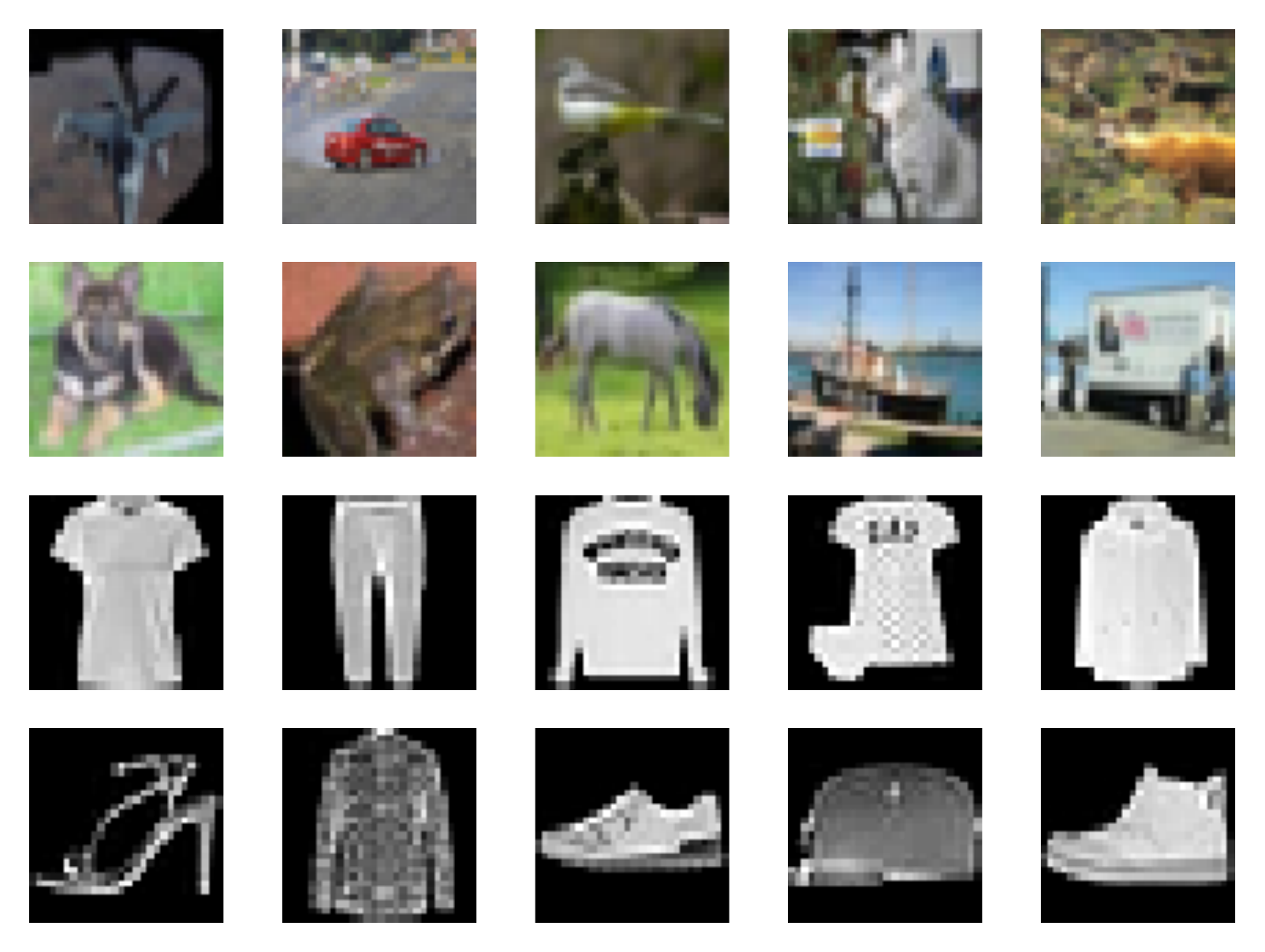}
    \caption{{\bf Samples of Image Datasets:} random samples from CIFAR-10 (color images) and from Fashion-MNIST (gray images).}
    \label{fig:image}
  \end{center}
\end{figure}

With the simulated MEC environment described above, we adopted two realistic object classification tasks using publicly-available large-scale image datasets. One was CIFAR-10, a classic object classification dataset consisting of 50,000 training images and 10,000 testing images with 10 object classes\footnote{\url{https://www.cs.toronto.edu/~kriz/cifar.html}}. This dataset has been used commonly in FL studies~\cite{Mcmahan2017,Konecny2016}. The other was Fashion MNIST~\cite{Xiao2017}, which comprised 60,000 training images and 10,000 testing images of 10 different fashion products such as T-shirts and bags. This dataset would give a more beneficial but sensitive setting because the ability to automatically recognize fashion products would be useful for various applications such as e-commerce, but the products that people are interested in are highly-private information. Figure~\ref{fig:image} shows sample images in the datasets.

For both tasks, the training dataset was distributed to $K=1000$ clients as follows: First, we randomly determined the number of image data owned by each client in a range of 100 to 1,000. Then, by following the experimental setup used in~\cite{Zhao2018}, we split the training dataset into the clients in two ways: \textbf{IID setting} where each client just sampled the specified number of images from the whole training dataset randomly, and \textbf{Non-IID setting} where each client sampled images randomly but from different subsets (2 out of the 10 categories chosen randomly) of the training data, standing for a more challenging but realistic setting. In each round of the FL protocols, we set $C=0.1$ based on \cite{Mcmahan2017} to select a maximum of $K\times C=100$ clients. Finally, the testing dataset was used only for measuring classification performances.

\subsection{Global Models and Their Updates}\label{sec:detail}
We implemented a standard convolutional neural network as a global model for both tasks. Specifically, our model consisted of six $3\times 3$ convolution layers (32, 32, 64, 64, 128, 128 channels, each of which was activated by ReLU and batch normalized, and every two of which were followed by $2\times 2$ max pooling) followed by three fully-connected layers (382 and 192 units with ReLU activation and another 10 units activated by soft-max). This resulted in approximately 4.6 million model parameters ($D\sub{m}=$ 18.3 megabytes in 32-bit float) for CIFAR-10 and 3.6 million parameters ($D\sub{m}=$ 14.4 megabytes in 32-bit float) for Fashion-MNIST. Deeper models such as residual networks~\cite{He2016} would provide higher classification performances. However, these models were not the focus of our experiments.

When updating global models, we selected the following hyperparameters according to \cite{Mcmahan2017}: $50$ for mini-batch size, $5$ for the number of epochs in each round, $0.25$ for the initial learning rate of stochastic gradient descent updates, and 0.99 for learning rate decay. The computation capability of each client was simply modeled by how many data samples it could process in a second to update a global model, which could be fluctuated due to other computation load on the client. We determined the mean capability of each client randomly from a range of 10 to 100, which are used the value for \texttt{Client Selection}. As a result, each update time, $t_k^{\rm UD}$, used in \texttt{Client Selection} varied from 5 to 500 seconds averagely. In \texttt{Scheduled Update and Upload}, the computation capability is determined by the Gaussian distribution with the standard deviation given by the $r\%$ of the mean capability value like our throughput model. We considered this range to be reasonable because our workstation required 5 seconds for a single update with a single GPU; mobile devices with a weaker computation resource could require a 10 or 100 times longer update time. Finally, we empirically set $T\sub{round}$ to 3 minutes and $T\sub{final}$ to 400 minutes. 

\subsection{Evaluation Details}

We compared \texttt{FedCS} with the FL protocol~\cite{Mcmahan2017} modified slightly to be limited with deadline $T\sub{round}$ for each round. We referred to this protocol as \texttt{FedLim}. In this baseline, the clients selected randomly by a MEC operator updated the models and sequentially uploaded their new parameters to a server until the deadline. The updates completed after the deadline were just discarded and not aggregated. \texttt{FedCS} and \texttt{FedLim} were evaluated based on the following metrics:
\begin{itemize}
\item {\bf Time of arrival at a desired accuracy (ToA@$x$)}: We observed the changes in the accuracy on testing datasets over time and identified when the accuracy reached a certain level for the first time (\ie, the earlier the better). Specifically, we report {\bf ToA@0.5} (\ie, 50\% accuracy) and {\bf ToA@0.75} for CIFAR-10 and {\bf ToA@0.5} and {\bf ToA@0.85} for Fashion-MNIST with the IID setting, and {\bf ToA@0.35} and {\bf ToA@0.5} for CIFAR-10 and {\bf ToA@0.5} and {\bf ToA@0.7} for Fashion-MNIST with the Non-IID setting.
\item {\bf Accuracy after the final deadline (Accuracy))}: We also measured the accuracy on testing datasets just after the final deadline ($T\sub{final}=360$ minutes since the beginning).
\end{itemize}

\subsection{Results}

\begin{table}[t]
\centering
\caption{{\bf Results obtained for CIFAR-10 and Fashion-MNIST with IID setting.} {\rm ToA@$x$: the time (in minutes) required to arrive at a testing classification accuracy of $x$ (the earlier the better). Accuracy: the testing accuracy after the final deadline. \texttt{FedLim} is an implementation of standard FL~\cite{Mcmahan2017} limited with the same deadline as that of \texttt{FedCS}. NaN means that the method did not achieve the required accuracy in some trials.}}
\begin{tabular}{@{}lccr@{}}
\toprule
\multirow{2}{*}{Method} & \multicolumn{3}{c}{CIFAR-10} \\
\cline{2-4}
& ToA@0.5 & ToA@0.75 & Accuracy \\
\midrule
\texttt{FedLim ($T_\mathrm{round}=3$ min)} & 38.1 & 209.2 & 0.77 \\
\midrule
{\bf \texttt{FedCS}} & & & \\
\, $T_\mathrm{round}=3$ min ($r = 0\%$) & \bf 25.8 & \bf 132.7 & \bf 0.79 \\
\, $T_\mathrm{round}=3$ min ($r = 10\%$) & \bf 27.9 & \bf 138.1 & \bf 0.78 \\
\, $T_\mathrm{round}=3$ min ($r = 20\%$) & \bf 31.1 & \bf 178.3 & \bf 0.78 \\
\, $T_\mathrm{round}=1$ min ($r = 0\%$) & NaN & NaN & 0.50\\
\, $T_\mathrm{round}=5$ min ($r = 0\%$) & 41.0 & 166.6 & 0.79\\
\, $T_\mathrm{round}=10$ min ($r = 0\%$) & 75.7 & 281.7 & 0.76\\
\midrule  
\midrule
\multirow{2}{*}{Method} & \multicolumn{3}{c}{Fashion-MNIST} \\
\cline{2-4}
& ToA@0.5 & ToA@0.85 & Accuracy \\
\midrule
\texttt{FedLim ($T_\mathrm{round}=3$ min)} & 10.4 & 66.8 & 0.90 \\
\midrule
{\bf \texttt{FedCS}} & & & \\
\, $T_\mathrm{round}=3$ min ($r = 0\%$) & \bf 10.6 & \bf 33.5 & \bf 0.91 \\
\, $T_\mathrm{round}=3$ min ($r = 10\%$) & \bf 11.3 & \bf 32.1 & \bf 0.92 \\
\, $T_\mathrm{round}=3$ min ($r = 20\%$) & \bf 12.7 & \bf 37.0 & \bf 0.91 \\
\, $T_\mathrm{round}=1$ min ($r = 0\%$) & 3.0 & 73.7 & 0.89\\
\, $T_\mathrm{round}=5$ min ($r = 0\%$) & 18.1 & 48.8 & 0.92\\
\, $T_\mathrm{round}=10$ min ($r = 0\%$) & 42.0 & 93.3 & 0.91\\
\bottomrule \\ 
\end{tabular}
\label{tab:result}
\end{table}



\textbf{IID setting: }
The main results with the IID setting are shown in Table~\ref{tab:result}. We ran each method ten times and computed the average ToA and accuracy scores. Overall, \texttt{FedCS} outperformed \texttt{FedLim} on both of the CIFAR-10 and Fashion-MNIST tasks in terms of ToA.
Specifically, \texttt{FedCS} achieved 75\% accuracy 76.5 minutes on average earlier than \texttt{FedLim} on CIFAR-10, and 85\% accuracy 33.3 minutes on average earlier on Fashion-MNIST when $T_\mathrm{round}=3$ and $r=0$. We also found that \texttt{FedCS} achieved a higher classification accuracy than \texttt{FedLim} after the final deadline (``Accuracy'' column in the table) than \texttt{FedLim} especially on CIFAR-10. These results indicate the improved efficiency of \texttt{FedCS} over \texttt{FedLim} in terms of the training progress. One reason for the improvement is because \texttt{FedCS} was able to incorporate much more clients into each training round: 7.7 clients for each \texttt{FedCS} while only 3.3 clients for \texttt{FedLim}, on average when $T_\mathrm{round}=3$.
Note that the current state-of-the-art accuracy is 0.9769 for CIFAR-10~\cite{Yamada2018} and 0.967 for Fashion-MNIST\footnote{\url{https://github.com/zalandoresearch/fashion-mnist}}. Nevertheless, our selection of model architectures was sufficient to show how our new protocol allowed for efficient training under resource-constrained settings and was not for achieving the best accuracies. The original FL~\cite{Mcmahan2017} without deadline limitations achieved accuracies of 0.80 for CIFAR-10 and 0.92 for Fashion-MNIST, both of which were comparable to the final performance of \texttt{FedCS}. We also confirmed that the uncertainty of throughput and computation capabilities, which were parameterized by $r$, did not greatly affect the performance of \texttt{FedCS}.


\begin{figure}[t]
  \begin{center}
    \includegraphics[width=\linewidth]{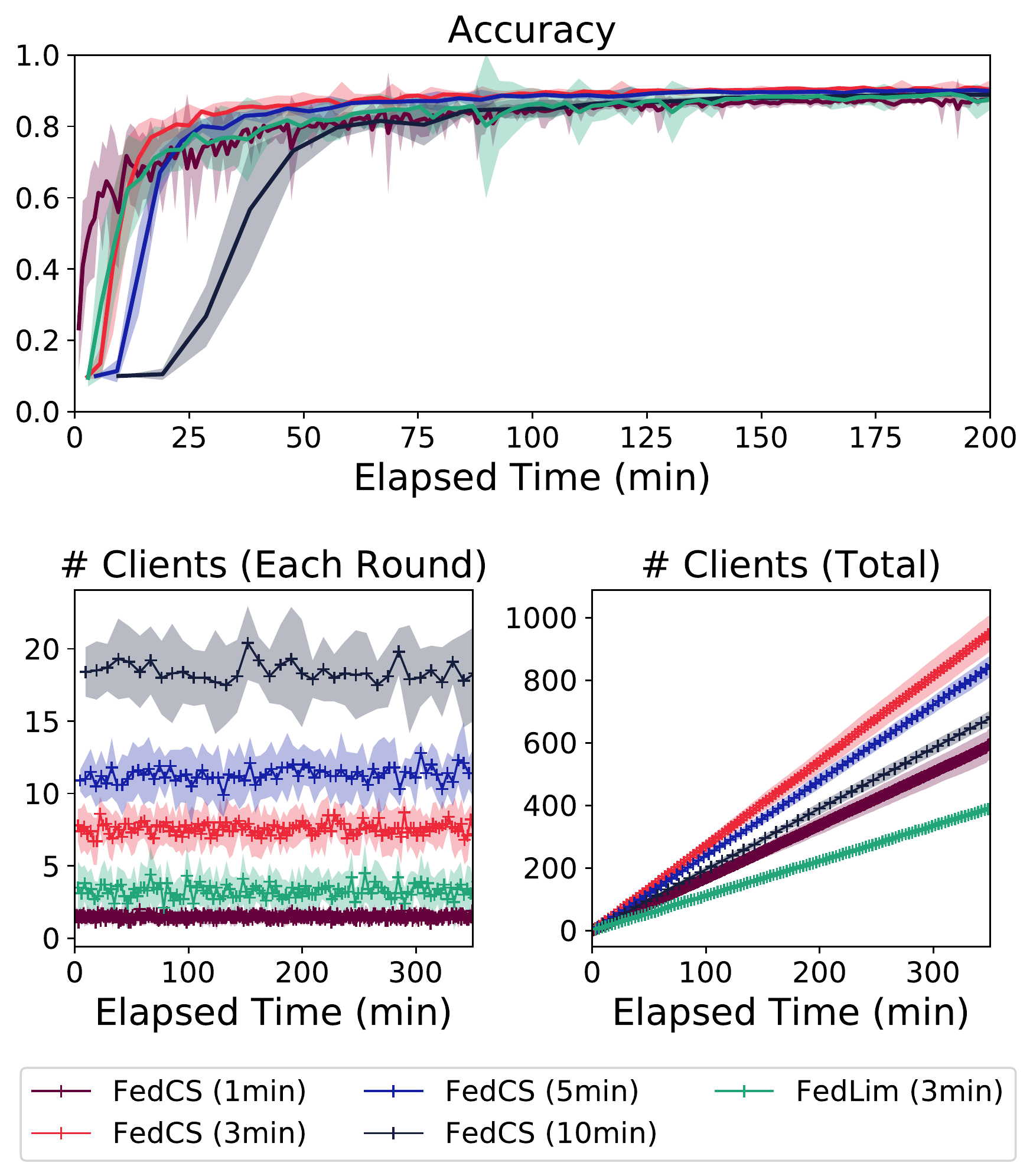}
    \caption{{\bf Effects of Different Values of Deadline $T\sub{round}$.} Top: accuracy curves; bottom-left: the number of clients selected in each round; bottom-right: the total number of selected clients. Shaded regions denote the standard deviation of the performance among ten trials.}
    \label{fig:client}
  \end{center}
\end{figure}

\textbf{Effect of $T\sub{round}$:} To obtain a deeper understanding of how our approach works, we investigated ToA and the changes in the classification accuracies \texttt{FedCS} on Fashion MNIST for different values of deadline $T\sub{round}$ while maintaining $T\sub{final}$ fixed, as shown in Table~\ref{tab:result} and Fig.~\ref{fig:client}. 
We observed that $T\sub{round}$ must be selected to be neither too long nor too short. While longer deadlines (\eg, 10 minutes) with \texttt{FedCS} involved numerous clients in each round, their performances were extremely limited owing to the smaller number of \texttt{Aggregation} steps. 
On the contrary, a short deadline, such as 1 minute, limited the number of clients accessible in each round, which also degraded the classification accuracies. 
A better method of selecting $T\sub{round}$ is to change it dynamically to involve a sufficient number of clients in each round. This is left for future work.

\begin{figure}[t]
  \begin{center}
    \includegraphics[width=\linewidth]{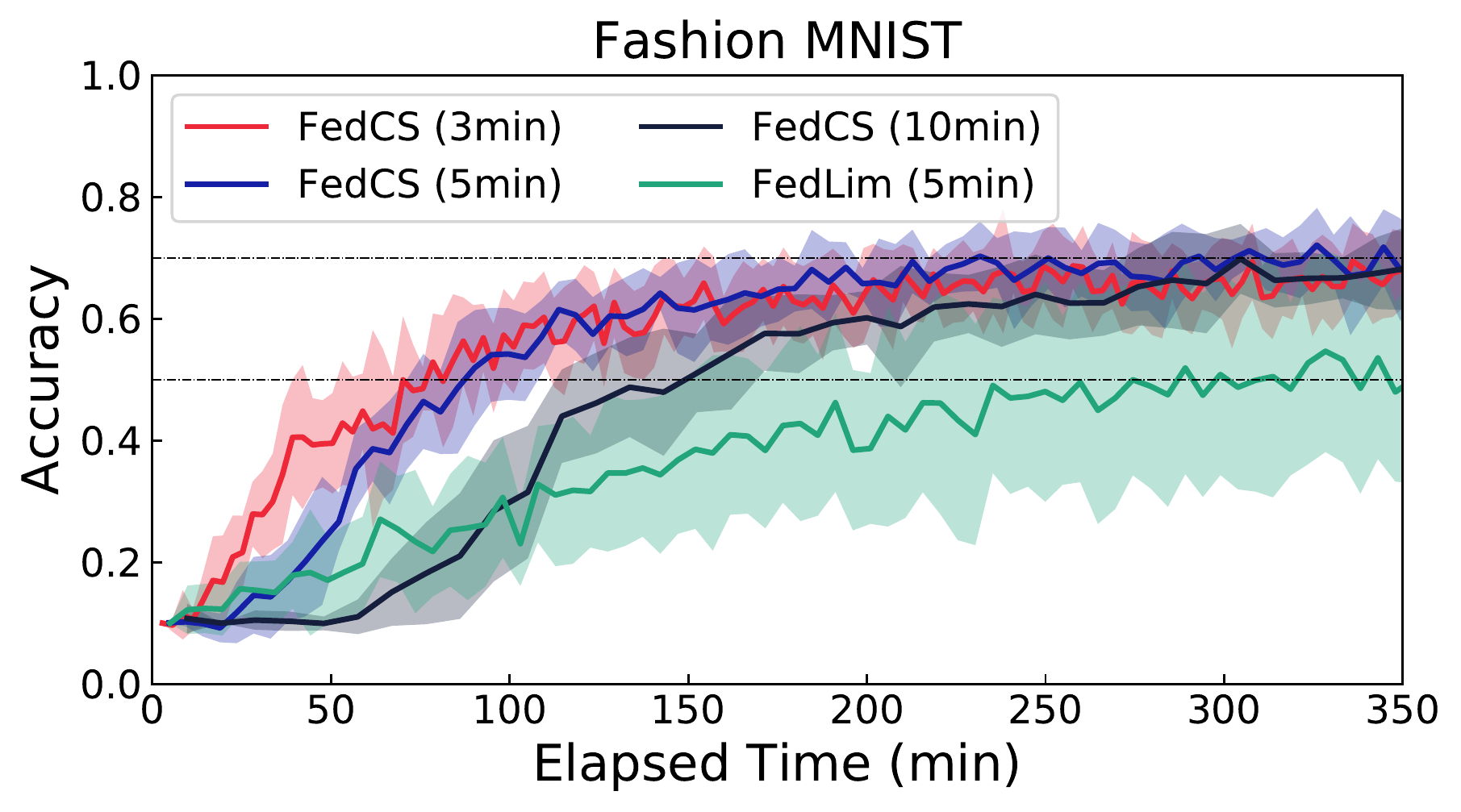}
    \caption{{\bf Accuracy curves of Different Values of Deadline $T\sub{round}$ in Non-IID setting.} Shaded regions denote the standard deviation of the performance among ten trials.}
    \label{fig:nonIID}
  \end{center}
\end{figure}

\textbf{Non-IID setting:} The results with the Non-IID setting are shown in Table~\ref{tab:result_nonIID} and Figure~\ref{fig:nonIID}. \texttt{FedCS} still works well while the performance of \texttt{FedLim} could not achieve the accuracy of even 50\% and 70\% on CIFAR-10 and Fashion-MNIST. However, similar to the previous work~\cite{Mcmahan2017}, the overall performances were limited with the Non-IID setting (\ie, lower averages and higher variances in the classification accuracies) compared to those with the IID setting. As indicated in the results of \cite{Mcmahan2017}, to better cope with non-IID data we need to increase either number of the selected clients for each round or that of rounds, both of which were however difficult due to the time constraints $T\sub{round}$ and $T\sub{final}$ we imposed in the experiments. One potential extension that can alleviate the non-IID problem is the additional use of model compression techniques~\cite{Konecny2016}, which could increase the number of clients that can be selected within the same constraint of $T\sub{round}$.


\begin{table}[t]
\centering
\caption{{\bf Results obtained for CIFAR-10 and Fashion-MNIST with Non-IID setting.} {\rm NaN means that the method did not achieve the required accuracy in some trials.}}
\begin{tabular}{@{}lccr@{}}
\toprule
\multirow{2}{*}{Method} & \multicolumn{3}{c}{CIFAR-10} \\
\cline{2-4}
& ToA@0.35 & ToA@0.5 & Accuracy \\
\midrule
\texttt{FedLim ($T_\mathrm{round}=5$ min)} & NaN & NaN & 0.31 \\
{\bf \texttt{FedCS}} ($T_\mathrm{round}=5$ min) & \bf 91.7 & \bf 213.7 & \bf 0.54 \\
\midrule  
\midrule
\multirow{2}{*}{Method} & \multicolumn{3}{c}{Fashion-MNIST} \\
\cline{2-4}
& ToA@0.5 & ToA@0.7 & Accuracy \\
\midrule
\texttt{FedLim ($T_\mathrm{round}=5$ min)} & NaN & NaN & 0.46 \\
{\bf \texttt{FedCS}} ($T_\mathrm{round}=5$ min) & \bf 82.4 & \bf 187.7 & \bf 0.71 \\
\bottomrule \\ 
\end{tabular}
\label{tab:result_nonIID}
\end{table}

\section{Conclusion}
\label{sec:concl}
We have presented a new protocol, \texttt{FedCS}, which aimed to perform FL efficiently in a MEC framework with heterogeneous clients. Our experimental results have revealed that \texttt{FedCS} constantly provided high-performance ML models in a significantly shorter time compared to the state-of-the-art protocol by incorporating more clients into its training process, regardless of the choices of datasets, the ways of splitting data (\ie, IID or Non-IID), and the uncertainty of throughput and computation capability.
As we limit our global model to sufficiently simple deep neural networks, other possible extension of this study is to train a more sophisticated model with dozens of millions of parameters using very large-scale data. 
Another interesting direction for future work is to work on more dynamic scenarios where the average amount of the resources as well as the required times for updating and uploading can fluctuate dynamically.


\section*{Acknowledgment}
This work was supported in part by JST ACT-I Grant Number JPMJPR17UK and JPMJPR16UT and KDDI Foundation.

\balance
\bibliographystyle{IEEEtran}
\bibliography{reference_short.bib}

\end{document}